\documentclass[twocolumn]{aastex631}
\usepackage{rotating}
\usepackage{graphicx}
\usepackage{xspace}
\usepackage{amsmath}	

\definecolor{myblue}{rgb}{0.13, 0.13, 0.55}

\newcommand{\halpha}          {H$\alpha$\xspace}
\newcommand{\kms}        {\ifmmode{\rm \,km\,s^{-1}}\else\,km\,s$^{-1}$\xspace\fi}
\newcommand{\lambdaobs}  {\ifmmode{\lambda_{\rm obs}}\else $\lambda_{\rm obs}$\xspace\fi}
\newcommand{\lambdaHa}  {\ifmmode{\lambda_{\rm H \alpha}}\else $\lambda_{\rm H \alpha}$\xspace\fi}

\shorttitle{Morpho-kinematics of NGC 6164/5}
\shortauthors{Lim et al.}
\graphicspath{{./}{figures/}}

\begin{document}

\title{A Morpho-Kinematic study of the Enigmatic Emission Nebula NGC 6164/5 Surrounding the Magnetic O-type Star HD 148937}

\correspondingauthor{Beomdu Lim}
\email{blim@kongju.ac.kr}

\author[0000-0001-5797-9828]{Beomdu Lim}\thanks{Sejong Science Fellow}
\affiliation{Department of Earth Science Education, Kongju National University, 
56 Gongjudaehak-ro, Gongju-si, Chungcheongnam-do 32588, Republic of Korea}
\affiliation{Earth Environment Research Center, Kongju National University, 56 Gongjudaehak-ro, Gongju-si, Chungcheongnam-do 32588, Republic of Korea}
\affiliation{Korea Astronomy and Space Science Institute, 776 Daedeok-daero, Yuseong-gu, Daejeon 34055, Republic of Korea}

\author[0000-0003-4071-9346]{Ya\"el Naz\'e}\thanks{Senior Research Associate FRS-FNRS (Belgium)}
\affiliation{Space Sciences, Technologies and Astrophysics Research Institute, Universit\'e de Li\`ege, Quartier Agora, All\'ee du 6 Ao\^ut 19c, B\^at. B5c, B-4000, Li\`ege, Belgium}

\author[0000-0002-0112-5900]{Seok-Jun Chang}
\affiliation{Max-Planck-Institut f\"{u}r Astrophysik, Karl-Schwarzschild-Stra$\beta$e 1, 85748 Garching b. M\"{u}nchen, Germany}

\author[0000-0003-4071-9346]{Damien Hutsem\'ekers}\thanks{Research Director FRS-FNRS (Belgium)}
\affiliation{Institut d’Astrophysique et de G\'eophysique, Universit\'e de Li\`ege, Quartier Agora, All\'ee du 6 Ao\^ut 19c, B\^at. B5c, B-4000, Li\`ege, Belgium}



\begin{abstract}
HD 148937 is a peculiar massive star (Of?p) with a strong magnetic field (1kG). The hourglass-shaped emission nebula NGC 6164/5 surrounds this star. 
This nebula is presumed to originate from episodic mass-loss events 
of the central O-type star, but the detailed formation mechanism is not yet well understood. Grasping its three-dimensional structure is essential to uncover the origin of this nebula. Here we report the high-resolution multi-object spectroscopic observation of NGC 6164/5 using the GIRAFFE on the 8.2m Very Large Telescope. Integrated intensity maps constructed from several spectral lines delineate well the overall shape of this nebula, such as the two bright lobes and the inner gas region. The position-velocity diagrams show that the two bright lobes are found to be redshifted and blueshifted, respectively, while the inner region has multiple layers. We consider a geometric model composed of a bilateral outflow harboring nitrogen-enriched knots and expanding inner shells. Its spectral features are then simulated by using a Monte-Carlo radiative transfer technique for different sets of velocities. Some position-velocity diagrams from simulations are very similar to the observed ones. According to the model that best reproduces the observational data, the two bright lobes and 
the nitrogen-enriched knots are moving away from HD 148937 at about 120 km s$^{-1}$. Their minimum kinematic age is 
estimated to be about 7,500 years. We discuss possible formation 
mechanisms of this nebula in the context of binary interaction. 
\end{abstract}



\section{Introduction} \label{sec:intro}
Massive stars are not only the sources of enormous ultraviolet radiation in the universe, but also the factories of heavy elements as they evolve. In addition, they are good tracers of star-forming regions in the galaxies \citep{GHV09} because their lifetime is only about several million years. In such regions, star formation may be regulated by feedback from massive stars in both constructive \citep{EL77} and destructive \citep{DEB12,DEB13} ways. At the end of stellar evolution, massive stars leave compact objects that can be associated with high-energy transient events such as $\gamma$-ray bursts. Hence, many fields of astronomy require an understanding of the formation and evolution of massive stars. In particular, mass loss is a crucial parameter to understand the evolution of massive stars \citep{S14}. 

HD\,148937 is a massive star of spectral type Of?p \citep{W72}. One specific feature of this spectral class is the presence of periodic line profile variations and those of HD\,148937 have the shortest period recorded up to now, only 7\,d \citep{NWM08}. As all other stars of this type, HD\,148937 has been found to be strongly magnetic, with a dipolar field of polar intensity 1\,kG \citep{WGG12}. The spectral peculiarities are then explained by an oblique magnetic rotator scenario \citep{BM97}. In this model, the stellar winds are channeled towards the equator by the strong field, where they collide. This confined material emits throughout the electromagnetic spectrum, with an enhanced and harder X-ray emission as well as numerous emission lines in the optical range. Because of the obliquity of the magnetic axis, the confined winds are seen under different angles throughout the rotation of the star, leading to the observed variations: the 7\,d period of HD\,148937 is thus interpreted as the rotation period of the star. However, the low amplitude of the variations, compared to other magnetic O-stars, suggested a low inclination for the star \citep{NUS10}, which was then confirmed by the magnetic monitoring \citep{WGG12}: HD\,148937 is thus seen nearly pole-on. In addition, interferometric and spectroscopic measurements recently indicated that HD\,148937 actually is a binary with two similarly massive components, a highly eccentric orbit, and a long orbital period \citep{WSE19}. Note that the inclination angle of the orbital plane of the system is also low. Finally, the stellar parameters of HD\, 148937 were well constrained by several previous studies (mass $\sim 50-60 M_\sun$, effective temperature $\sim 40000$K, $\log g \sim 4.0$, $v\sin i \geq 45$ km s$^{-1}$ -- \citealt{NWM08,NUS10,WGG12,MHB15}). It is also worth noting that HD 148937 shows a nitrogen enrichment as high as seen in O-type supergiants \citep{MHB15}.

HD\,148937 is surrounded by several layers of nebulosities. \citet{H59} was the first to recognize that the nebulae NGC\,6164 and NGC\,6165 actually appear symmetrically with respect to the star. He then proposed the two features to belong to one single elongated structure of about 5$^{\prime}$ diameter, at the time identified as a planetary nebula. In addition, a series of arcs delineate a larger shell of about 12$^{\prime}$ radius centered on the star and even larger nebular features can be spotted at a distance of 64$^{\prime}$ from the star \citep{W61}. The largest structure is interpreted as a Str\"omgren sphere linked to the intense ultraviolet (UV) radiation of the massive star \citep{FPH85}, the broken-arc shell as a bubble blown by the strong stellar wind of the massive star(s), and the NGC\,6164/5 pair as recent ejecta \citep{LC87}. While ejected material is commonly spotted around evolved massive stars (of Wolf-Rayet or luminous blue variable type), NGC\,6164/5 is a rare example of nebulosities associated to a ``normal", unevolved massive star, making its study crucial to better understand the mechanisms of stellar feedback. 

The kinematics of the NGC\,6164/5 nebula has been studied several times in the past. First, the overall difference between the two parts was recognized: NGC\,6164, to the NW of the star, displays positive radial velocities (RVs) while NGC\,6165, to the SE of the star, shows negative RVs \citep{CF70}. This is characteristic of material expanding from a center, so it was taken as an additional argument in favor of the NGC\,6164/5 unification scheme. More detailed investigations, however, revealed the complexities of the kinematic structure, with velocities ranging from +117 to --70\,km\,s$^{-1}$ in NGC\,6164 and from --170 to --3\,km\,s$^{-1}$ in NGC\,6165 \citep{P74,CA86}. Even nearby areas could show very different velocities \citep{P74}. In fact, the nebular line profiles often reveal several components, especially in the inner parts of the nebula, demonstrating the presence of several layers along the line-of-sight \citep{LC87}. Some previous studies suggested that the main lobes may be part of either a spiral structure \citep{MHN17}, a bipolar structure \citep{P74,LC87}, or a helical structure \citep{CA86}.

The abundances of NGC\,6164/5 were also studied in detail. \citet{DM77} already mentioned an enhanced He abundance while the detailed study of \citet{DPH88} found an enrichment in N and He as well as a depletion in Ne and O (see also \citealt{LC87}). These non-solar abundances constitute an additional argument in favor of a nebula formed by an ejection event from HD\,148937. It should be noted in this context that the star itself displays a clear nitrogen enrichment, as would be expected for such a scenario. However, in a counter-intuitive way, the enrichment was found to be larger in the farther lobes than in the inner parts of the nebula \citep{DPH88}. This was confirmed in the Herschel data analysis of \citet{MHN17}. The inner parts close to the star (structure `H1' in that paper) displayed a N/O ratio about one dex above the solar abundance while the C/O ratio appeared three times the solar one. Larger values were found for the outer NW lobe (structure `H2'), which also appeared denser. Assuming the nebular material was ejected by the star, its abundances would then reflect those at the surface of the star at the time of the ejection. Compared with predictions of single-star evolution models, the observed nebular abundances then suggest an ejection about 0.6\,Myr ago for the lobes and about 1\,Myr ago for the inner parts. 

Note that the extinction was often found to be rather uniform across the nebula (e.g. \citealt{LC87}) although a dark cloud has been reported in the area \citep{PF09}. 
Interpreting the various nebular features in a coherent scheme has proven to be difficult. Assuming a single event, one could think of ejection following a merger. This scenario is linked to the fact that HD\,148937 is magnetic and to the theoretical proposal that magnetic fields could be generated in merging events (\citealt{TWL08,FPT09} - see also review by \citealt{L12}). However, it may be difficult to reach a rotation as slow as that of HD148937 so soon after the merging. Besides, the detection of a companion by \citet{WSE19} potentially implies that the history of the system is more complex and it remains to be demonstrated that the initial triple system can keep the distant companion in the currently observed orbit after the merger. 

Because of the bipolar appearance of the nebula, its axis, assumed to be similar to the stellar (rotation) or orbital axis, was often thought to be close to the plane of the sky, i.e. the star would be seen equator-on. In this case, the inner parts would correspond to slower material ejected first in the equatorial plane while the distant lobes would correspond to a second event having faster ejecta collimated along the polar axis. This configuration would be similar to what is seen in the famous $\eta$\,Carinae homunculus \citep{S08}. However, the inclinations of the rotation and orbital axes were clearly determined to be low, i.e. the system is rather seen pole-on. In such a case, one could instead consider the inner parts to come from a polar ejection and the lobes from a subsequent equatorial ejection. It may be noted that \citet{DM77} had already proposed two different directions of ejection for the inner parts and the lobes, with a possible pole-on configuration. However, the inner parts should then present the largest red- or blue-shifted velocities, which is not observed \citep{P74,CA86,LC87}. Besides, the multiple velocity components are not easily reconciled with a simple ejection event as envisaged in such scenarios. 

Based on the above arguments, a more complex configuration needs to be anticipated. \citet{CA86} has notably proposed a helical geometry, where the north and south ejection flows could get into the same line-of-sight, as well as different parts of the same flows. However, it is unclear how the material would flow along the polar axis away from the star. Indeed, magnetic channeling works in the other direction. Definitely, a more in-depth investigation, with modern means, of NGC\,6164/5 is warranted, with the hope of clarifying the exact geometry of the ejected flows. 

We perform the spectroscopic study of NGC 6164/5 with a high spectral resolution and a dense fiber configuration. The aim of this study is to uncover the three-dimensional (3D) geometry of the nebula and investigate its origin in the context of massive star evolution. In Section 2, the observational data we used are presented. The observational features of NGC 6164/5 are investigated using integrated intensity maps and position-velocity (PV) maps in Section 3. We simulate a model based on the 3D distribution of the nebula in position-velocity space and present a comparison of the observational data with the simulated one in Section 4. The formation of this nebula is discussed in Section 5, and our results are summarized in Section 6.


\begin{figure}[t]
\epsscale{1.0}
\plotone{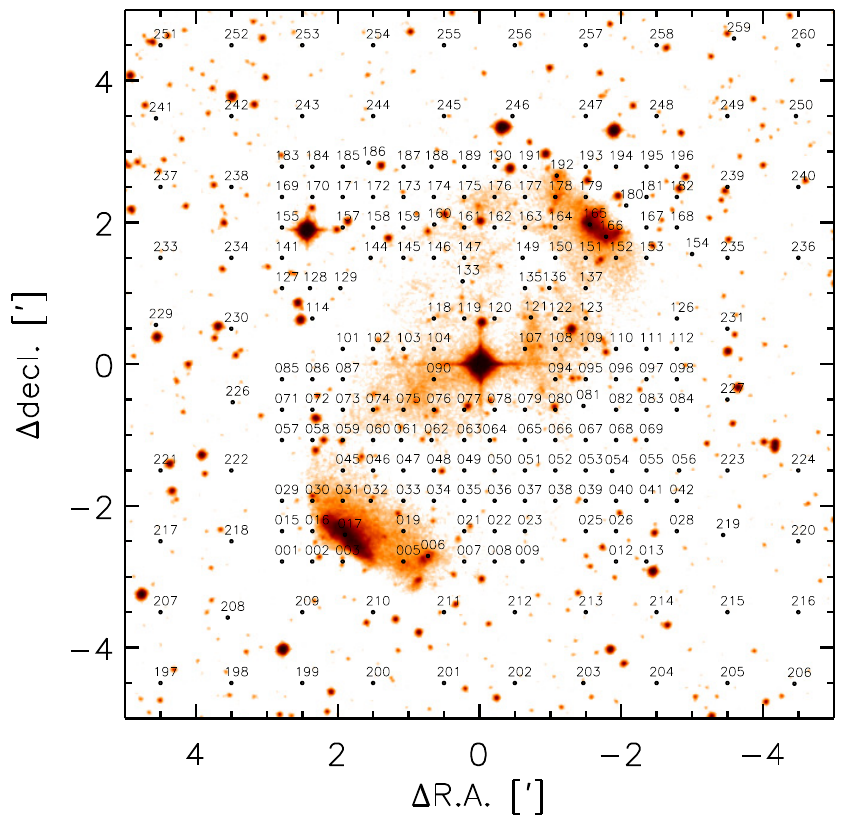}
\caption{Positions of the medusa fibers on the Digitized Sky Survey 
image \citep{RBB91,L94,DGO98}. This optical image can be found in MAST: \url{http://dx.doi.org/10.17909/T9QP4J}. The fiber position numbers are labeled on the image. The positions of the fibers are relative to HD 148937 
(R.A. = $16^{\mathrm{h}} 33^{\mathrm{m}} 52\fs387$, decl. = $-48^{\circ} 06^{\prime} 40\farcs477$).}\label{fig1}
\end{figure}

\begin{figure}[t]
\epsscale{1.0}
\plotone{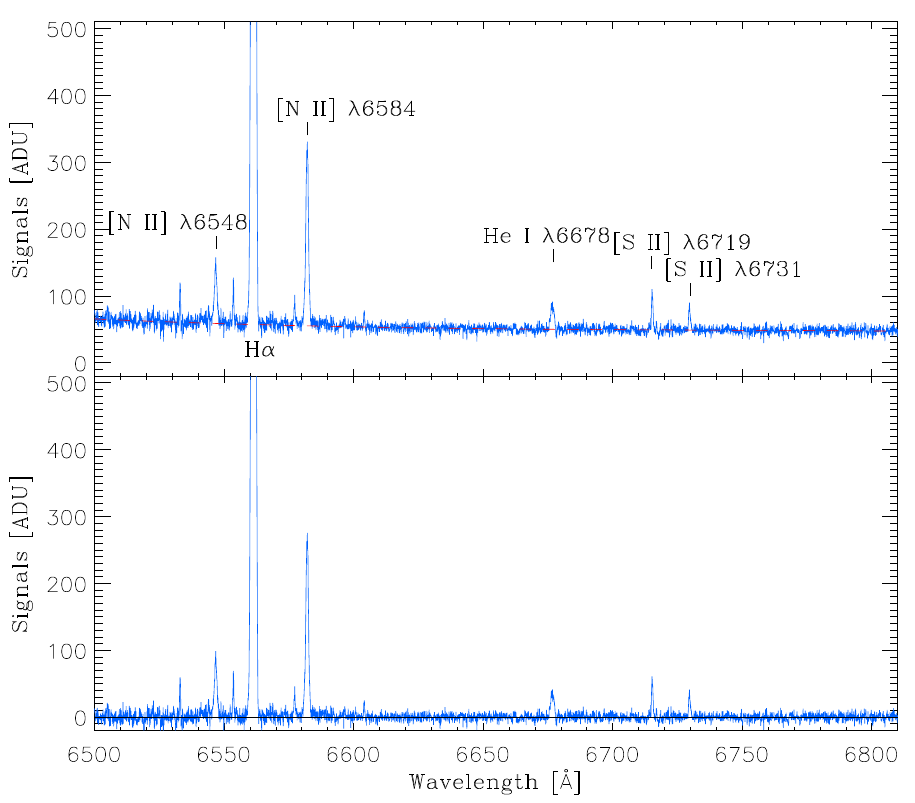}
\caption{One-dimensional spectrum of a given position on NGC 6164/5. 
In the upper panel, the red dashed line traces the continuum level 
from the third-degree polynomial fitting, while the lower panel shows 
the continuum-subtracted spectrum. Nebular spectral lines are labeled 
in the upper panel.}\label{fig2}
\end{figure}

\section{Observations and Data} \label{sec:floats}
The spectroscopic observations of NGC 6164/5 were performed with 
the European Southern Observatory 8.2m Very Large Telescope 
(UT2) on Cerro Paranal in Chile on 2019 March 24. We took a total of 
220 spectra over the nebula using the intermediate to high-resolution 
spectrograph GIRAFFE\footnote{\url{https://www.eso.org/sci/facilities/paranal/instruments/flames}}, whose fibers are fed by FLAMES \citep{PAB02}. A total of three frames were obtained for the same 
field. The exposure time of each frame was set to 1195s. 
Figure~\ref{fig1} displays the positions of 
the FLAMES fibers . The order-separating filter HR15N was used to 
cover the spectral range of 6470 \AA \ to 6790 \AA. Some dome flat 
and ThAr lamp spectra were also obtained after the target 
observation for calibration. 

The observational data were reduced by following the standard 
reduction procedures under the EsoReflex environment \citep{FRB13}. 
We extracted one-dimensional spectra using the {\tt IRAF/SPECRED} 
package. Figure~\ref{fig2} exhibits one of the extracted spectra. 
Polynomial fitting was performed on the continua of all 
spectra (see the red dashed line in the upper panel), and the 
continua were subtracted by the best-fit curves to obtain 
usable normalized spectra. Our spectra show the usual
nebular emission lines in this domain, e.g., [N {\scriptsize II}] $\lambda\lambda$6548, 
6584, H$\alpha$, He {\scriptsize I} $\lambda6678$, and 
[S {\scriptsize II}] $\lambda\lambda6719, 6731$.

We fit the emission lines H$\alpha$, [N {\scriptsize II}] 
$\lambda$6584, and [S {\scriptsize II}] $\lambda\lambda$6716, 
6730 with Gaussian profiles using the {\tt MIDAS} command 
DEBLEND/LINE. Since most lines have several velocity components along 
the line of sight, multiple Gaussian profiles were used to fit 
such complex line profiles. The RVs were taken from the centers of 
the best-fit Gaussians and then converted to heliocentric RVs. We 
also obtained the full width at half maximum and fluxes of the 
lines.

\begin{figure*}[t]
\epsscale{1.2}
\plotone{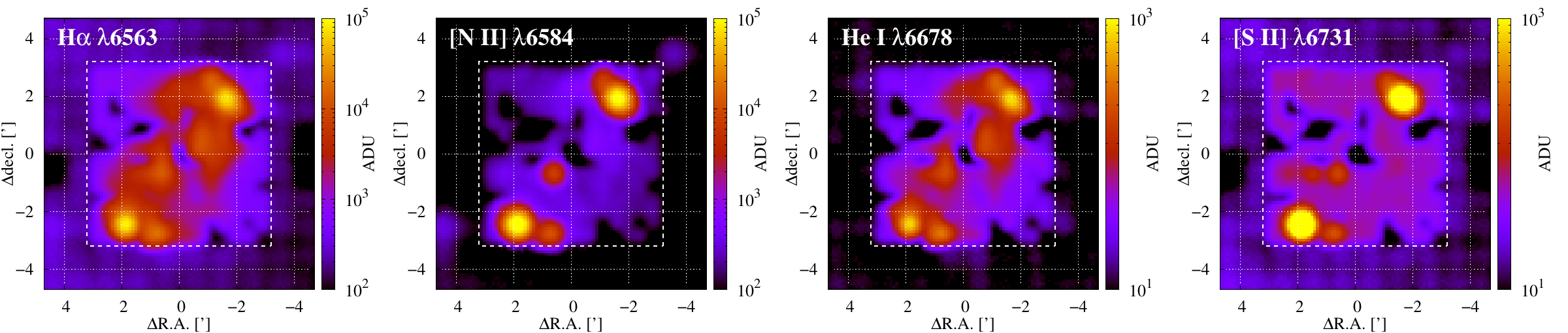}
\caption{Integrated intensity maps from four spectral lines: H$\alpha$ 
(first panel), [N {\scriptsize II}] $\lambda$6584 (second panel), He {\scriptsize I} 
$\lambda$6678 (third panel), and [S {\scriptsize II}] $\lambda$6731 (fourth panel). 
The signals in the RV range from $-200$ km s$^{-1}$ to 200 km s$^{-1}$ are 
integrated at a given fiber position. The integrated intensities are shown 
in a logarithmic scale. The boxes outlined by dashed lines represent a 
$6^{\prime} \times 6^{\prime}$ region from HD 148937. These maps are smoothed by 
using a Gaussian smoothing technique, where the kernel sizes of $0\farcm25$ and 
$0\farcm5$ are adopted for the inner ($6^{\prime} \times 6^{\prime}$) 
and outer regions, respectively. A clear difference can be observed 
between permitted  (H$\alpha$ and He {\scriptsize I} 
$\lambda$6678) and forbidden ([N {\scriptsize II}] $\lambda$6584 and [S {\scriptsize II}] $\lambda$6731) lines.}\label{fig3}
\end{figure*}

\section{Results}
\subsection{Integrated Intensity Maps}
Integrated intensity maps from different emission lines are 
useful tools to investigate the spatial distribution of the gas. 
We created integrated intensity maps using a Gaussian smoothing 
technique for the positions and counts (ADU) from given fiber 
positions. Figure~\ref{fig3} displays the integrated intensity 
maps from the four different emission lines H$\alpha$, 
[N {\scriptsize II}] $\lambda$6584, He {\scriptsize I} 
$\lambda$6678, and [S {\scriptsize II}] $\lambda$6731. 

The integrated intensity map of H$\alpha$ line 
reproduces well the overall structure seen in optical images. 
The diffuse gas is detected mostly over NGC 6164/5. 
The region extending out of the nebula (outside of the 
white box delineated by a dashed line in Figure~\ref{fig3}) 
may be part of the filamentary halo heated by shocks and 
photoionization from the central O-type star \citep{FPH85}. 
The northern and southern lobes contain bright knots. 
The He {\scriptsize I} $\lambda$6678 line traces a 
similar structure for the nebula. However, this line 
was not detected in the halo.

The [N {\scriptsize II}] $\lambda$6584 map highlights 
the presence of the northern and southern lobes, while 
the line strength of the diffuse gas is not as strong 
as those seen in the H$\alpha$ map. The 
[S {\scriptsize II}] $\lambda$6731 map traces 
almost the same features as those in the [N {\scriptsize II}] 
$\lambda$6584 map, and the halo gas was also detected. 
Given the fact that the formation of these two forbidden lines 
is sensitive to electron density, the two main lobes 
have different electron densities (or density structures) 
from the other regions.

\begin{figure}[t]
\epsscale{1.1}
\plotone{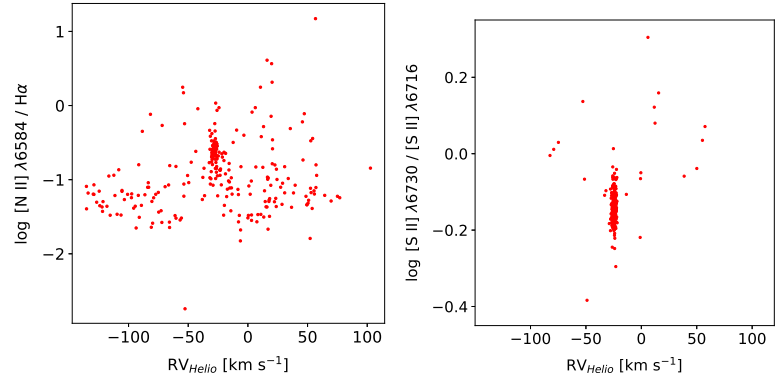}
\caption{Distributions of $\log$[N {\scriptsize II}] $\lambda$6584/H$\alpha$ (left panel) 
and $\log$[S {\scriptsize II}] $\lambda$6730 / [S {\scriptsize II}] 
$\lambda$6716 (right panel) with respect to RVs. The integrated flux of 
individual lines were obtained from the best-fit Gaussian distributions.}\label{fig4}
\end{figure}

The spectral lines of NGC 6164/5 are affected 
by emission from the filamentary halo \citep{FPH85} 
as seen in the integrated intensity map for H$\alpha$ 
lines. Figure~\ref{fig4} displays the line ratios ($\log$[N{\scriptsize II}] 
$\lambda$6584/H$\alpha$ and $\log$[S {\scriptsize II}] 
$\lambda$6730/[S {\scriptsize II}]) with respect to RVs. 
Some emission components are centered in a narrow RV range between 
$-18$ km s$^{-1}$ and $-32$ km s$^{-1}$ around the system 
velocity of HD 148937, and they show moderate spreads in the 
two line ratios ($-1.3$ to $0.0$ for $\log$[N {\scriptsize II}] $\lambda$6584/
H$\alpha$ and $-0.4$ to 0.3 for $\log$[S {\scriptsize II}] $\lambda$6730/
[S {\scriptsize II}] $\lambda$6716). Most of these emissions may originate from the filamentary 
halo. Meanwhile, the other components associated with NGC 6164/5 
show a much broader ranges of RVs and of line ratios. In order to 
probe the kinematics of NGC 6164/5, it is necessary to minimize 
the contribution of the halo component. The emission components 
associated with the halo were discarded from further analysis.

\begin{figure*}
\begin{center}
\includegraphics[width=16cm]{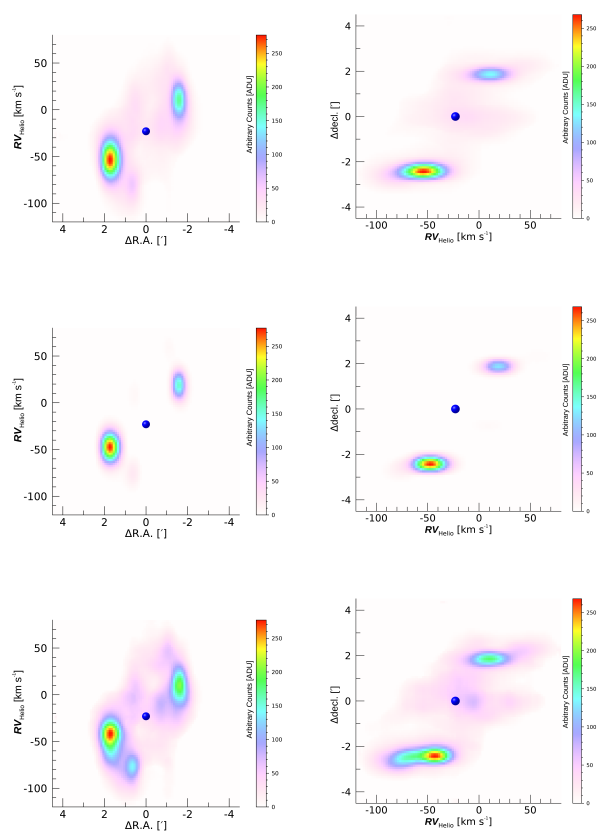}
\end{center}
\caption{Position-velocity diagrams obtained from H$\alpha$ (upper panel), [N 
{\scriptsize II}] $\lambda$6584 (middle panel), and He {\scriptsize I} $\lambda$6678 
(lower panel). The blue sphere represents the central O-type star HD 148937. The contour 
shows the integrated intensities in ADU. The coordinates are relative to HD 148937 
as shown in Figure~\ref{fig1}}\label{fig5}
\end{figure*}

\begin{figure*}
\begin{center}
\includegraphics[width=16.0cm]{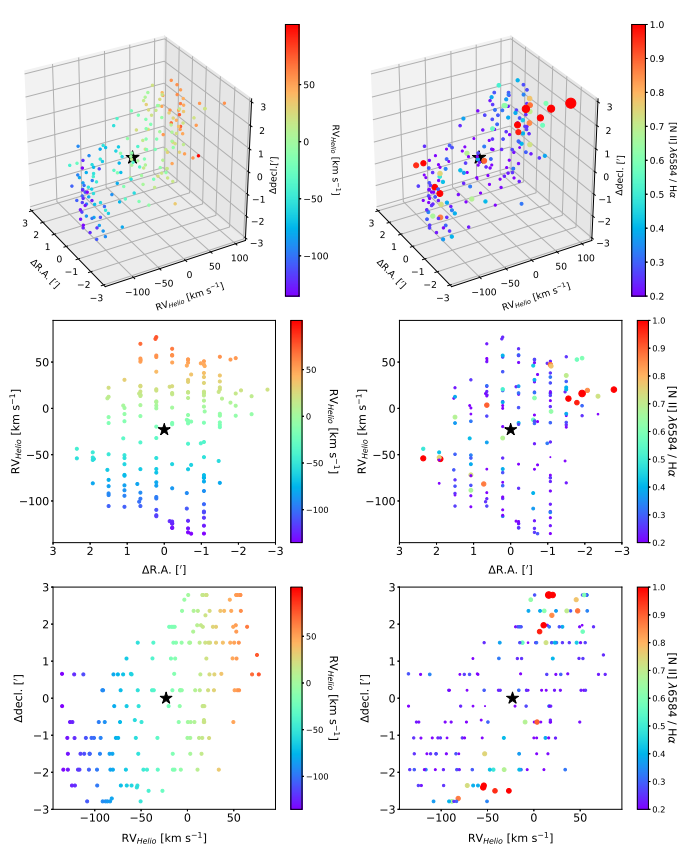} 
\end{center}
\caption{PV diagrams (left panels) and [N {\scriptsize II}] 
$\lambda$6584/ H$\alpha$ line ratios (right panels) of 
each gas component. In the left panels, 
the RVs of individual gas components are shown 
by different colors. The symbol sizes and colors are 
proportional to the line ratios [N {\scriptsize II}] 
$\lambda$6584/ H$\alpha$ in the right panels. The star symbols 
represent HD 148937.}\label{fig6}
\end{figure*}

\subsection{PV diagrams}
We investigated the velocity field of NGC 6164/65 in 
position-velocity (PV) space. Our multi-object spectroscopic 
data contain four-dimensional information, R.A., decl., 
velocity, and intensity. However, the fibers were 
not placed in a regular grid. A Delauney triangulation 
technique was applied to interpolate the intensities 
and velocities of the emission lines (H$\alpha$, 
[N {\scriptsize II}] $\lambda$6584, and He {\scriptsize I} 
$\lambda$ 6678) into data cubes composed of $90\times90\times90$ 
regular volume cells \citep{LSB18,LNG19,LNH21}. We 
then constructed the PV diagrams by the sum 
of intensities along R.A. and decl., respectively. 

Figure~\ref{fig5} displays the PV 
diagrams of NGC 6164/5. We also plotted a dot to mark 
the position of the central O star HD 148937 in the 
diagrams. This star is known 
to be a binary \citep{WSE19}, and its systemic RV is 
about $-23$ km s$^{-1}$ \citep{WSE19}. In the PV diagrams, 
there are two prominent gas components with respect to 
HD 148937, corresponding to the northern and southern 
lobes. The knots in the northern and southern lobes are 
moving away from the central star at RVs of about 33 km s$^{-1}$ 
and $-27$ km s$^{-1}$, respectively, i.e. they correspond to a 
symmetrically expanding feature. 

H$\alpha$ and He {\scriptsize I} $\lambda$6678 lines 
trace the distribution of hot gas filling the space 
between the two lobes, while [N {\scriptsize II}] 
$\lambda$6584 line is too weak to probe the velocity 
field of the inner hot gas. The RVs of the inner 
hot gas from H$\alpha$ and He {\scriptsize I} $\lambda$6678 
lines ranges from $-75$ km s$^{-1}$ to 50 km s$^{-1}$, 
equivalent to $-52$ km s$^{-1}$ and 73 km s$^{-1}$ 
relative to HD 148937, respectively. 

The spatial resolution of our multi-object spectroscopy 
is about 0\farcm5$\times$0\farcm5 for the inner 
6\farcm0$\times$6\farcm0 region, which is higher 
than in previous spectroscopic observations. Despite this, 
some small structures can be blurred 
due to the limited resolution and uneven grid of fibers. In 
particular, velocity components with weak line intensities 
may be more affected by such effects in the construction 
of PV diagrams. We thus investigated the individual 
velocity components decomposed from the best-fit multiple 
Gaussian distributions.

Figure~\ref{fig6} displays the 3D PV diagrams. The 
RV distribution of individual gas components are 
consistent with that shown in Figure~\ref{fig5}. 
Also, we confirmed that there are high-velocity 
gas components between the two main lobes. Their 
velocities relative to the central O-type star 
exceed 100 km s$^{-1}$, which is faster than the 
velocities of the two main lobes. The PV 
diagrams in Figures~\ref{fig5} and ~\ref{fig6} 
show an almost symmetric velocity field, indicating 
that the nebula surrounding HD 148937 may have a 
axisymmetric 3D structure.

The [N {\scriptsize II}] $\lambda$6584/H$\alpha$ 
line ratios of the individual components are shown 
by different sizes of dots in the right panels 
of Figure~\ref{fig6}. The majority of points with 
high [N {\scriptsize II}] $\lambda$6584/H$\alpha$ 
line ratios (red dots) are found in the knots of each lobe 
although some are close to the central region 
as seen in Figure~\ref{fig3}. The envelope of the 
two lobes and the inner hot gas tend to have 
lower [N {\scriptsize II}] $\lambda$6584/H$\alpha$ 
line ratios.

\section{Model}

\subsection{Geometry}
The morphology of NGC 6164/5 has been explained 
by three different geometric models invoking a helical structure 
\citep{CA86}, a spiral structure \citep{MHN17}, 
and a bipolar structure \citep{P74,LC87}. In this study, 
we revisited the geometric model of the bipolar 
structure. Since HD 148937 has a pole-on configuration 
\citep{NUS10,WGG12}, we will consider that NGC 6164/5 was probably 
ejected from the equator of the star. Hereafter, the bipolar 
structure is thus referred to as the bilateral structure.

If the main lobes from the plane of the sky were not inclined from 
the plane of the sky, its RVs would not be detected by spectroscopy. This 
inclination angle is one important parameter 
in setting up a geometric model. In order to infer the inclination 
angle, it is necessary to constrain the velocity of the 
nebula in the actual (physical), 3D space, from the known RVs. 
The diagrams shown in Figures~\ref{fig5} and ~\ref{fig6} indicate that some 
parts of the H$\alpha$ emitting nebula have RVs exceeding 100 km s$^{-1}$ 
relative to the central star. Such high-RVs may be close to 
actual velocities if the associated nebular region is moving 
nearly parallel to the line of sight but may be very different 
from the actual ones if inclined on the line of sight.

\begin{figure}
\includegraphics[width=7cm]{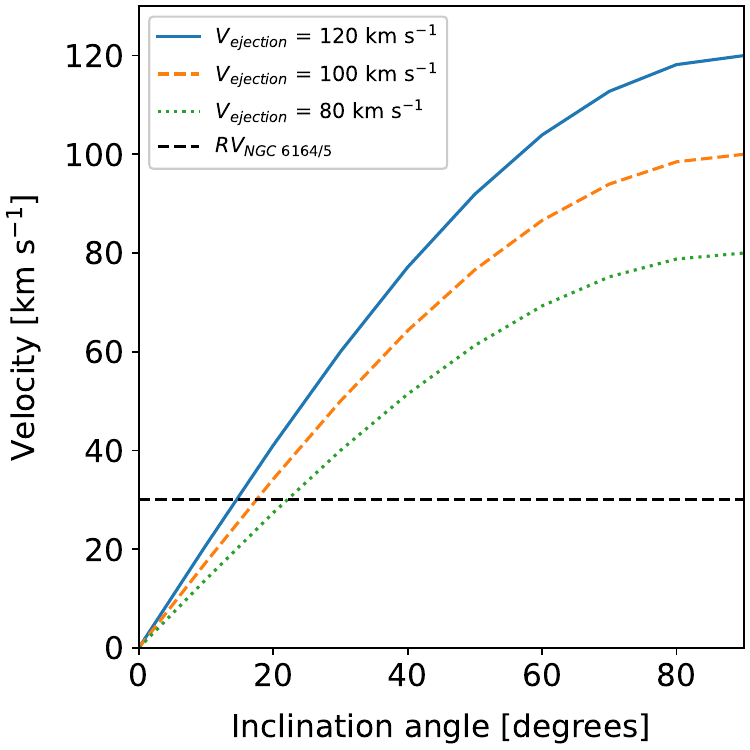}
\caption{Expected velocities of ejecta with respect to different 
inclination angles. We considered ejection velocities of 80 
(dotted line), 100 (dashed line), and 
120 (solid line) km s$^{-1}$. The observed {\it RV}, measured on the main lobes, 
is shown by a dashed line.}\label{fig7}
\end{figure}

\begin{figure*}
\begin{center}
\includegraphics[width=16cm]{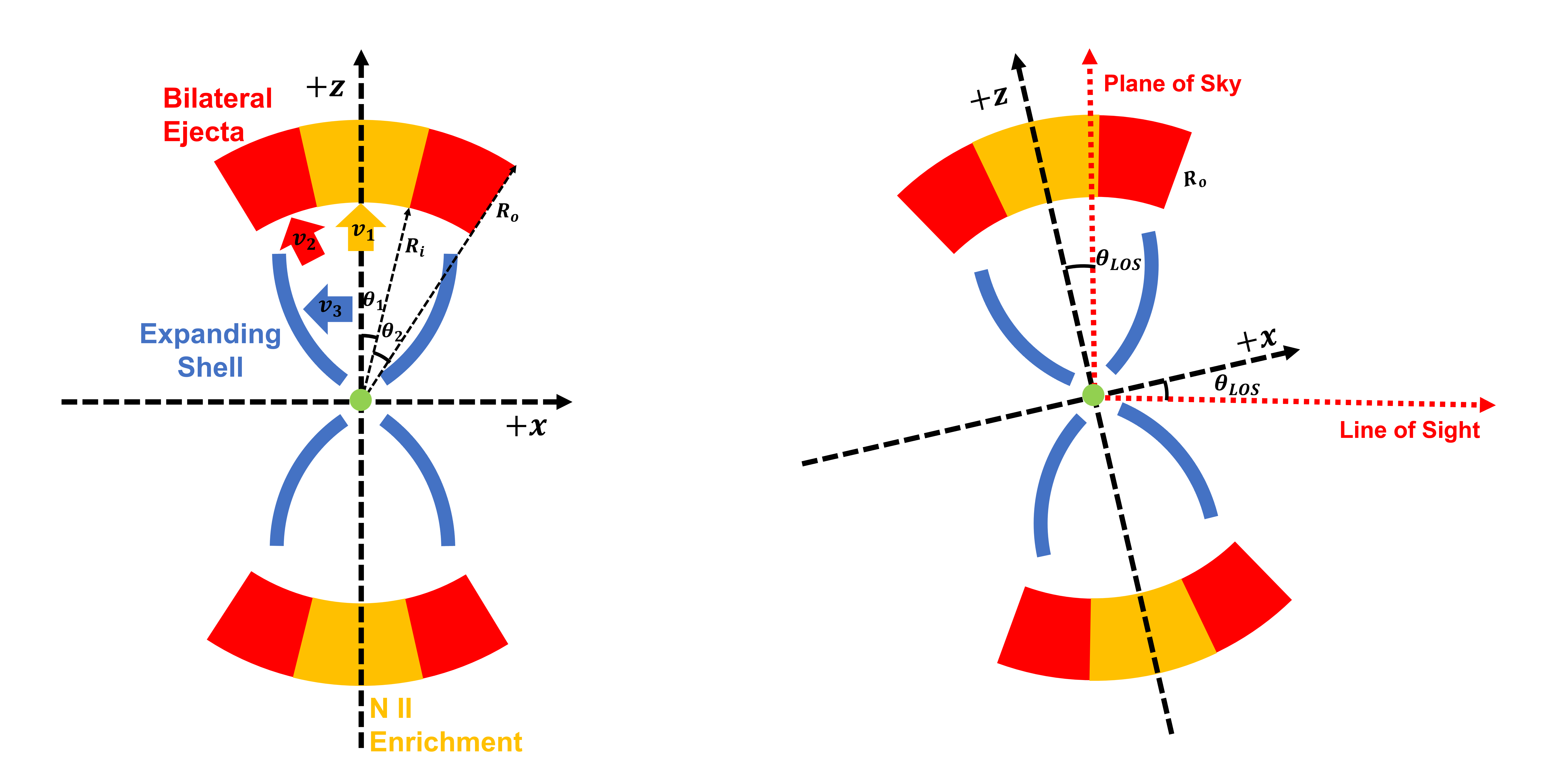}
\end{center}
\caption{Schematic illustration of our bilateral model in a cartesian 
coordinate. The $x$-axis corresponds to the rotational axis, and 
the equator of the central star is placed along the $y$-$z$ plane. 
The left cartoon displays the structure and velocity vectors. The 
nebula has three components, nitrogen-enriched knots (orange), envelopes 
(red), and expanding hollow shells (blue). The velocities of the knots 
and envelopes ($v_1$ and $v_2$) are proportional to the distance from the central 
star (green dots). The hollow shells are also expanding from $z$-axis at $v_3$. 
The right cartoon shows the inclination of this system 
from the line of sight or the plane of sky. The inclination angle 
($\theta_{LOS}$) of 20$^{\circ}$ is adopted in this model 
(see the main text for detail). 
}\label{fig8}
\end{figure*}

We thus computed the expected RVs of the main lobes with respect 
to different inclination angles for three ejection velocities 
(80, 100, and 120 km s$^{-1}$) as shown in Figure~\ref{fig7}. 
Given the fact that the RVs of knots in the main lobes are 
about 30 km s$^{-1}$ relative to the central star with an 
uncertainty of several km s$^{-1}$, the inclination angle 
can then be constrained in a range of 
15$^{\circ}$ to 22$^{\circ}$ (see the dashed line in the 
figure). In what follows, we thus adopted an inclination angle 
of 20$^{\circ}$. This is larger than that the 5$^{\circ}$ angle 
inferred by \citet{LC87} but it remains compatible with the rotational 
inclination angle ($\leq 30^{\circ}$) measured from the line 
of sight \citep{WGG12}, i.e. the angle between the plane of sky 
and the equator of the O-type star. 

\subsection{Setup}\label{sec:setup}
Figure~\ref{fig8} displays the schematic illustration 
of our bilateral model. The $x$-axis corresponds to the 
rotational axis of HD 148937, and the $y$-$z$ plane is thus 
an equatorial plane. The rotational axis is inclined from 
the line of sight by $\theta_{\mathrm{LOS}}$ (see right 
cartoon of Figure~\ref{fig8}). In this study, $\theta_{\mathrm{LOS}}$ 
was set to 20$^{\circ}$. 

We considered that NGC 6164/5 is composed of three components: 
nitrogen-enriched knots (orange), extended envelopes (red) 
surrounding the nitrogen-enriched knots and expanding thin 
(hollow) shells. The thickness of the knots and 
envelopes are characterized by an outer radius $R_o$ and inner 
radius $R_i = 0.7 R_o$. Both opening angles of the knots and envelopes, 
$\theta_1$ and $\theta_2$, were set to 15$^\circ$. The hollow shells 
occupy an inner region characterized by the inner and outer 
radii decreasing as a function of $\cos {\theta_z}$, where $\theta_z$ 
is measured from $z$-axis and ranges from $\theta_1 + \theta_2$ to 
75$^{\circ}$. Therefore, the inner radii decrease from $0.4R_{o}\cos \theta_z$ to $\sim 0.10 R_{o}$ (close to the central star), and the outer ones 
decrease $0.5R_{o} \cos \theta_z$ to $\sim 0.13 R_{o}$. 

The knots display N [{\scriptsize II}] $\lambda$6584/H$\alpha$ ratios 
higher than those of the envelopes. Their similar position 
may indicate that the knots have almost the same expanding 
velocities as the envelopes or slightly higher velocities. We 
therefore postulated that the knots and envelopes may have 
different origins.

The velocities of knots and envelopes were set to increase with the distance from 
the central star and reach $v_1$ (knots) and $v_2$ (envelopes) 
at $R_o$. This study considered the situation that the two 
components are comoving ($v_1 = v_2$). Two different setups 
for their velocities at $R_o$, 100 km s$^{-1}$ and 120 km 
s$^{-1}$ as shown in Figure~\ref{fig7}, were 
applied to our simulations. On the other 
hand, the hollow shells are inflated from $z$-axis at a constant 
velocity $v_3$. Three different setups for $v_3$ 
(80, 100, and 120 km s$^{-1}$) were considered in our simulations.

\subsection{Simulation}

We developed a new 3D Monte Carlo radiative transfer simulation 
to investigate our spectral cube data, based on the Monte-Carlo 
techniques used in \cite{CL20} \footnote{The 
Monte-Carlo radiative transfer simulation in \cite{CL20} is 
developed to study scattering processes with atom hydrogen. We refer 
to it to use basic Monte-Carlo techniques in our simulation.}.
In this simulation, we set the kinematics and geometry of the emission 
nebula as described in Section~\ref{sec:setup}.
In order to compare our model with observational data, our simulations 
uniformly generate a total of $10^6$ photon packets
within the model nebulae. Each photon packet carries a velocity  $v_{\rm kine}$ and a position ${\bf r_i} = (x_i,\ y_i,\ z_i)$ at emitting location. Furthermore, these components have different intensities for \halpha and [N {\scriptsize II}] $\lambda$6584 lines (see Figure~\ref{fig3}), and therefore the weight factors of the photon packets, representing relative emissivity in a unit volume, of the nitrogen-enriched knots, envelopes, and hollow shells were set to 3, 1, and 0.1, respectively. After that, we collected photons along the line of sight. Since the scope of our study is to explore the kinematic 
structure of NGC 6164/5 in comparison with the observed data, 
detailed physical parameters such as electron densities and the number 
of photons from the central star were not considered in these simulations.

The model nebula in 3D coordinates is projected onto the 
plane corresponding to the plane of the sky. The number of photons 
at any given point was summed along the line of sight to construct 
the integrated intensity map. Given that the unit vector of 
the line of sight ${\bf \hat{k}}$ defined by
\begin{equation}
{\bf \hat{k}_p} = (\sin\theta_p \cos\phi_p,\ \sin\theta_p \sin\phi_p,\ \cos\theta_p),
\end{equation}
\noindent where $\theta_p$ and $\phi_p$ are the azimuthal and polar angles of the line of sight. The basis vectors of the projected plane ${\bf \hat{x}_p}$ and ${\bf \hat{y}_p}$ 
are given by 
\begin{eqnarray}
{\bf \hat{x}_p} &=& (\sin\phi_p,\ -\cos\phi_p,\ 0), \\ \nonumber
{\bf \hat{y}_p} &=& (-\cos\phi_p \cos\theta_p,\ -\sin\phi_p \cos\theta_p,\ \sin\theta_p),
\end{eqnarray}
$\phi_p$ is fixed at $0^\circ$ because the geometry is symmetric to $z-$axis.
$\theta_p$ is $20^\circ$ because it corresponds to $\theta_{\rm LOS}$.

The positions and RVs in the projected plane are needed 
to construct the PV diagram as shown in Figure~\ref{fig6}.
The coordinates in the projected plane are newly defined using the projected coordinates $x_p$ and $y_p$, which correspond to R.A. and decl., respectively, 
given by
\begin{equation}
x_p = {\bf r_i} \cdot {\bf \hat{x}_p}, \quad y_p = {\bf r_i} \cdot {\bf \hat{y}_p}.
\end{equation}

The Doppler factor $\Delta V$ of the collected photon, which is equivalent to the RV, is given by
\begin{equation}\label{eq:velocity}
    \Delta V = - {\bf v_{\rm kine}} \cdot {\bf \hat{k}},
\end{equation}
where  $\bf v_{\rm kine}$ is the velocity at the location emitting the photon, which is computed by $v_1$, $v_2$, and $v_3$.

\begin{figure*}
\includegraphics[width=16cm]{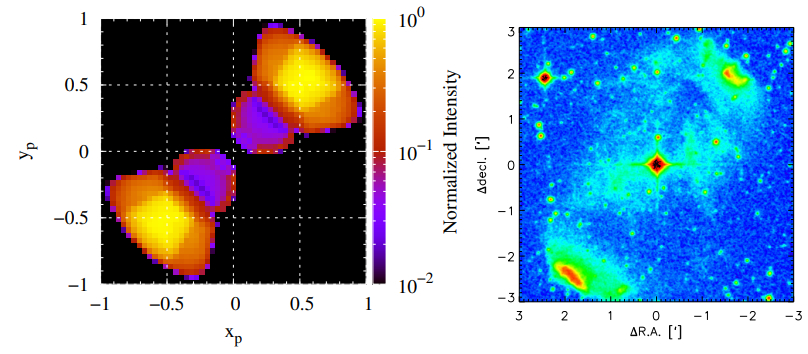}
\caption{Integrated intensity map of our bilateral model (left panel) and the Digitized 
Sky Survey image (\citealt{RBB91,L94,DGO98}, right panel). The left panel shows the intensity map in the projected radius, considering the weight factors of 3, 1, and 0.1 for the nitrogen-enriched knots, envelopes, and hollow shells, respectively. Thus, the bright regions to the north and south are composed primarily of photons emitted from nitrogen-enriched knots. The color bar represents the level of normalized intensity. The Cartesian coordinates $x_p$ and $y_p$ 
correspond to R.A. (different sign) and Dec., respectively. The right panel displays 
the optical image for comparison with our model. The color map highlights the structure 
of NGC 6164/5.}\label{fig9}
\end{figure*}

In our simulation, we applied the projection effects to the 
the structure and velocities of the modelled nebula. The following 
section will discuss the comparison with the simulated and observational data via the integrated intensity map (Figure~\ref{fig9}) and the PV diagram (Figure~\ref{fig10}) .

\subsection{Comparison with the observational data}

The integrated intensity map was rotated by $45^\circ$ 
clockwise in the $x_p-y_p$ plane to reproduce the tilt of NGC 6164/5 as seen 
in Figure~\ref{fig1}. Figure~\ref{fig9} shows comparison of 
the integrated intensity map of our model with the 
observational image from Digitized Sky Survey \citep{RBB91,L94,DGO98}. The 
observational image shows intensity variation across 
the nebula because of thin dust lanes. Such dark regions 
are often found in H {\scriptsize II} regions and 
are not very bright in infrared passbands. Their presence 
can only be recognized with bright background illumination. 
In our simulations, we did not consider the presence of thin dust 
lanes in front of NGC 6164/5.

\begin{figure*}
\includegraphics[width=\textwidth]{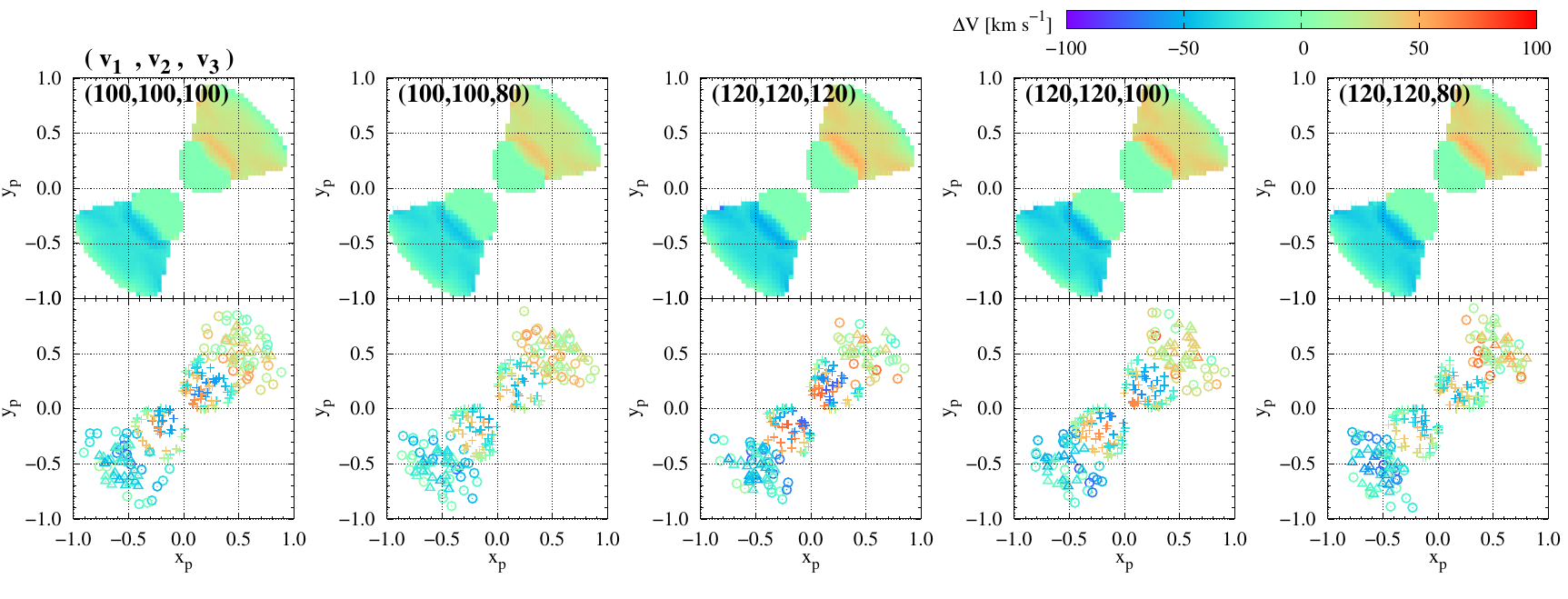}
\caption{Maps of mean RVs (top panels) and RVs of 
random spots (bottom panels) for different setups, 
calculated from the Doppler shifts of the expanding 
substructures. In the bottom panels, nitrogen-enriched knots, 
envelopes, and hollow shells are shown by triangle, circle, 
and plus symbols, respectively.}\label{fig10}
\end{figure*}

We first investigated the global velocity trends. To this aim, 
we averaged along the line of sight the Doppler shifts of the 
photons (upper panels of Figure~\ref{fig10}). The northern and southern 
lobes have, on average, positive and negative RVs, respectively. 
Meanwhile, the mean velocities of the hollow shells are nearly 0 km s$^{-1}$ 
because multiple components along the line of sight include both 
backward- and forward-moving gas. This is reminiscent of what is found 
in observation (Figure~\ref{fig5}).

In order to probe the local variations of the RVs, the Doppler 
shifts were calculated at about 200 arbitrary spots within the 
3D structure. The lower panels of Figure~\ref{fig10} exhibit the positions 
of such spots projected onto the sky, with different symbols 
depending on their associated substructure (knots, lobes, or shell) 
and different colors depending on their RVs (red/blue for 
forward/backward-moving gas). As can be seen, there are multiple 
components with different RVs along the line of sight in the 
inner hollow shells and envelopes, as in observations.

\begin{figure*}
\includegraphics[width=\textwidth]{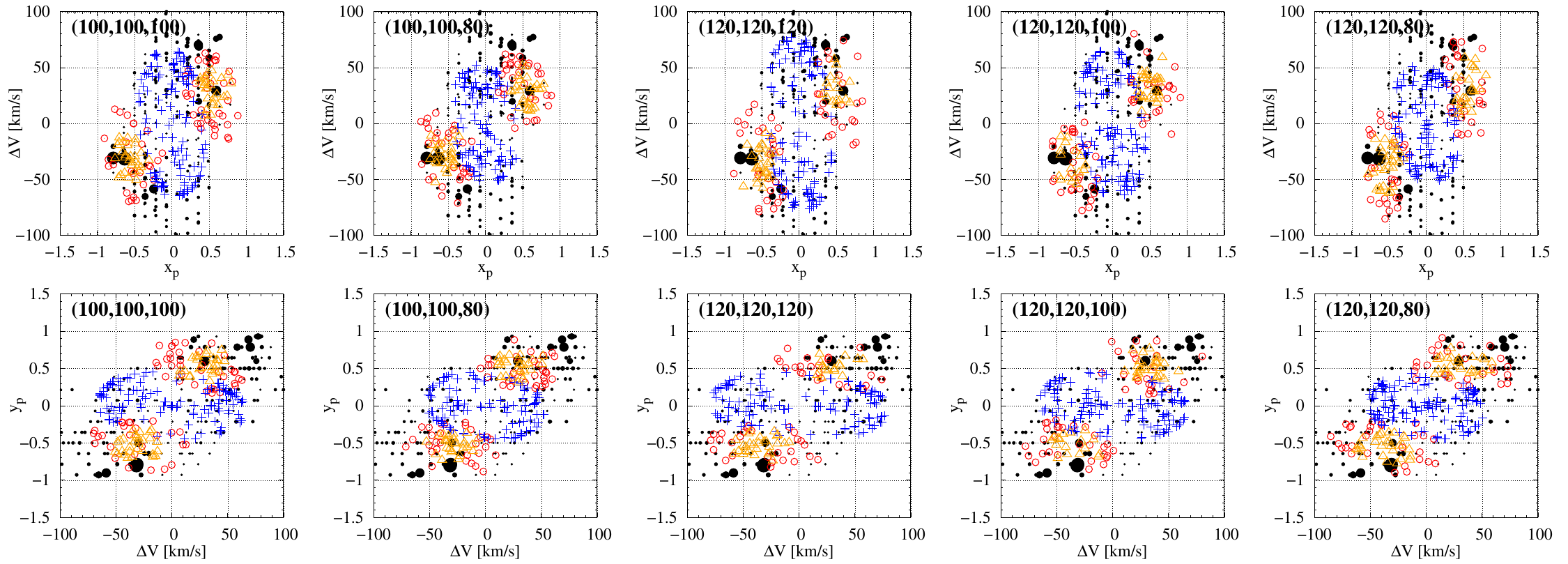}
\caption{PV diagrams of the model nebula along $x_p$ (upper panels) 
and $y_p$ (lower panels). The choices of $v_1$, $v_2$, and $v_3$ are 
provided in the upper left corner of each panel. The symbols are the 
same as those in the lower panel of Figure~\ref{fig10}, and their 
colors correspond to those of three gas components in the schematic 
illustration of Figure~\ref{fig8}. The filled circles represent the 
observed PV diagrams from H$\alpha$ 
lines shown in Figure~\ref{fig6}, and their size is proportional to the line ratio 
[N {\scriptsize II}] $\lambda$6584 / H$\alpha$.}\label{fig11}
\end{figure*}

Figure~\ref{fig11} shows the PV diagrams taken from the randomly 
sampled data in Figure~\ref{fig10}. In the simulated PV diagrams, 
photon-emitting components can be traced, as shown by different 
symbols. In all simulations, the nitrogen-enriched knots 
are found at around $\pm30$ km s$^{-1}$ as found in observations. 
On the other hand, the inner hollow shells cover different RVs, 
depending on $v_3 = 80$, 100, and 120 km s$^{-1}$. 

We directly compared the simulated PV diagrams 
with the observed ones (black dots) in Figure~\ref{fig11}. 
The model nebulae occupy a region covering a range from 
$-1$ to $+1$ for the $x_p$ and $y_p$ axes (arbitrary units). 
Our spectroscopic observation covers the overall 
structure of NGC 6164/5 which corresponds to a (true) size of 
($6^{\prime} \times 6^{\prime}$). For comparison, the 
observed PV diagrams were scaled by dividing the position 
relative to the star by 3 in arcminutes. Compared to Figure~\ref{fig6}, note also 
that the observed RV were corrected by the stellar RV ($-23$ 
\kms). The models adopting low velocities ($v_1=v_2=$ 100 km s$^{-1}$, 
$v_3=$ 80 or 100 km s$^{-1}$) tend to underestimate RVs of 
the nebula. On the other hand, the PV diagrams from the models 
assuming high-velocities agree best with the observed ones, 
particularly for $v_1=v_2=v_3 =120$ km $s^{-1}$. 

\section{Discussion}
\citet{LC87} considered that HD 148937 has an equator-on 
configuration, and they modeled the mass ejection along the 
rotational axis of HD 148937 which they considered as nearly perpendicular to the line 
of sight. However, both spectroscopic and magnetic monitoring 
observations support a pole-on configuration 
\citep{NWM08,WGG12}. Based on these more recent observations, 
a new simulation of the geometry of NGC 6164/5 was performed in this study. 
Our simulation reproduced well the observed velocity field, 
particularly for the kinematics of the inner region. It is 
now useful to compare the properties of NGC 6164/5 with 
those of mass ejection events from other massive stars 
in order to understand the possible formation mechanism of the emission nebula.

In the 19th century, $\eta$ Car has ejected a large amount of mass 
(10--35$M_{\sun}$) at velocities of about 650 km s$^{-1}$ 
\citep{SF07,MSG16}. Some ejecta of the Homunculus 
nebula are moving at velocities of several hundred to 
several thousand km s$^{-1}$ \citep{WDB01,S08}. NGC 6164/5 
may not originate from such a $\eta$ Car-like giant eruption 
in terms of mass and velocity. Indeed, HD 148937 has lost some 
2 $M_{\sun}$ \citep{MHN17} and the ionized gas is receding 
away at maximum 100 km s$^{-1}$. In addition, there is 
a significant difference between the Homunculus nebula and 
NGC 6164/5. The great eruption of $\eta$ Car occurred along 
the polar axis.

The emission nebula M 1-67 surrounding the Wolf-Rayet star 
WR124 is composed of a bipolar structure and a torus 
\citep{ZTS22}. The bipolar component has an expanding 
velocity of 88 km s$^{-1}$ \citep{SNP98}, while the  
velocities of the torus range from 30 km s$^{-1}$ 
to 60 km s$^{-1}$ \citep{ZTS22}. It is often thought that 
WR 124 does not have a companion, however, 
several observational studies do not definitely rule out 
this possibility \citep{TOH18,MLS82,GKF10}. \citet{ZTS22} 
explained the formation of M 1-67 and WR124 in the context 
of the common envelope scenario \citep{P67}.

According to \citet{WSE19}, HD 148937 has a massive 
O-type companion in a very eccentric orbit ($e = 0.75$)  
with an orbital period of 26 years. In that study, the 
inclination of the orbital plane from 
the plane of the sky was deduced to be about 40$^{\circ}$. If the 
plane where NGC 6164/5 is expanding is nearly parallel to 
the orbital plane, the origin of the nebula could be associated 
with a binary interaction. Unlike the M 1-67 system, 
NGC 6164/5 may not be formed through the common envelope scenario 
because the eccentric orbit and the long orbital period prohibit 
the two stars to interact closely as in a common envelope scenario. 
However, other binary interactions due to tidal force could 
possibly occur when the primary and secondary stars come close 
together at periastron. This periastron passage can therefore 
result in mass ejection events for highly eccentric binaries 
\citep{RBM05,MKH11}. Such a process could also explain why ejection 
occurs in a specific direction of the equatorial plane, rather than 
over the full equator.

The knots in the main lobes are about $2\farcm7$ away 
from HD 148937. Since the distance to the central star 
is about 1.1 kpc according to the offset-corrected 
Gaia parallax \citep{gdr3,LBB21}, the projected distance 
of the knots is about 0.86 pc. If we consider the inclination 
angle of 20$^{\circ}$, the actual distance between the 
central star and the knots is about 0.92 pc. With a velocity of 120 km s$^{-1}$, 
the travel time of the knots from the central star to their current positions 
is estimated to be about 7,500 years assuming 
that the knots have rapidly accelerated and reached their final 
velocity on a short timescale. If a gradual acceleration of the knot 
is considered, the kinematic age would be somewhat longer than 7,500 
years. This kinematic age is much smaller than 0.6 - 1.3 Myr, 
the age estimated from the comparison of the N/O abundances with 
evolutionary models \citep{MHN17}.

Here, we have shown that we can reproduce the appearance 
and kinematics of the nebula by assuming that all gas components were ejected 
at almost the same velocity. If the nitrogen-enriched knots 
were ejected at the same time at slower velocities, these 
knots would not reach their current position. The overall 
morphology of NGC 6164/5 would then be different from 
the current one. Therefore, these knots should have at 
least velocities similar to the other nebular components 
with less nitrogen enrichment if ejected simultaneously.

The puzzling difference in nitrogen abundance between 
the different features may then be understood if their 
origins are different. For example, knots and other features 
could be associated with the primary and companion stars, 
respectively. Given the fact that the nitrogen abundance of 
HD 148937 is as high as those of O-type supergiants \citep{MHB15}, 
the primary star (49 $M_{\sun}$ -- \citealt{WSE19}) may be evolving 
into the supergiant stage, while its companion star (34 $M_{\sun}$) 
may still be on main-sequence stage. 

$\eta$ Car and several eccentric binaries are often assumed to have 
undergone eruptions around periastron passages 
\citep{vGS07}. The brightenings seen in light curves 
were interpreted as the result of the deformation 
of the stellar surfaces by tidal forces on the primary 
star. While the tidal force can strongly affect the stellar surface 
of binary pairs, it is still not fully understood how the binary 
interaction can/could drive a mass ejection. Nevertheless, 
we propose the origin of NGC 6164/5 to be similar. 
In order to confirm this idea, it is necessary to monitor 
the light curve of HD 148937 on a long timescale and 
spectroscopically find the signature of other mass ejection 
events around periastron passages. In addition, 
theoretical approaches utilizing smoothed particle 
hydrodynamics will give us a better understanding of 
mass ejection processes in detail.

\section{Summary}
The magnetic O-type star HD 148937 is surrounded by the emission nebula 
NGC 6164/5. In this study, we investigated 
the kinematic and morphological properties of the nebula with high-resolution 
spectroscopy. 

We measured six strong emission lines ([N {\scriptsize II}] 
$\lambda\lambda$6548, 6584, H$\alpha$, He {\scriptsize I} 
$\lambda$6678, and [S {\scriptsize II}] $\lambda\lambda$6719, 6731) 
in our data. The integrated intensity maps 
constructed from H$\alpha$ and He {\scriptsize I} $\lambda$6678 
lines show the overall structure of the nebula, while the two 
forbidden lines [N {\scriptsize II}] $\lambda$6584 and 
[S {\scriptsize II}] $\lambda$6731 only trace the bright lobes.

The morpho-kinematics of NGC 6164/5 was investigated in PV 
diagrams. It confirms that the bright lobes are receding 
away from HD 148937 at RVs of about 30 km s$^{-1}$. The bright 
knots in these two lobes show high [N {\scriptsize II}] 
$\lambda$6584 / H$\alpha$ line ratios indicative 
of nitrogen enrichment. In addition, some nitrogen-enriched 
spots were also detected closer to HD 148937. 

On the other hand, the gas surrounding 
the main lobes and in the inner regions tends to have lower 
[N {\scriptsize II]} $\lambda$6584 / H$\alpha$ ratios. 
In those regions, multiple components at different RVs are detected, indicating 
the presence of multiple layers along the line of sight. 

In order to interpret the observed PV diagrams, we 
constructed a geometric model to reproduce the spectroscopic results. 
This model assumes a hollow expanding shell close to the star and 
lobes further away. The envelopes of the lobes host nitrogen-enriched knots. 
Overall, the modeled nebula is expanding along the equatorial plane, which 
is tilted by 20$^{\circ}$ from the plane of the sky. 
To match the data, all features should expand with $\sim$ 120 \kms. 
This yields a kinematic age of 7500 years. A single ejection velocity thus seems 
sufficient to reproduce the observed features, but abundances differ between them. 
This could however be explained in a binary interaction model, with ejection occurring 
at periastron from both stars. Further observational 
and theoretical studies are required to understand 
the mass ejection process of this system in detail.\\

\noindent The authors thank the anonymous referee for constructive 
comments and suggestions. This work presents results from 
the European Space Agency (ESA) space mission Gaia. Gaia data 
are being processed by the Gaia Data Processing and Analysis 
Consortium (DPAC). Funding for the DPAC is provided by national 
institutions, in particular the institutions participating in the 
Gaia MultiLateral Agreement (MLA). The Gaia mission website 
is {\url https://www.cosmos.esa.int/gaia}. The Gaia archive website 
is {\url https://archives.esac.esa.int/gaia}. The Digitized Sky Surveys 
were produced at the Space Telescope Science Institute under U.S. Government grant NAG W-2166. The images of these surveys are based on photographic data obtained using the Oschin Schmidt Telescope on Palomar Mountain and the UK Schmidt Telescope. The plates were processed into the present compressed digital form with the permission of these institutions. The Second Palomar Observatory Sky Survey (POSS-II) was made by the California Institute of Technology with funds from the National Science Foundation, the National Geographic Society, the Sloan Foundation, the Samuel Oschin Foundation, and the Eastman Kodak Corporation. The Oschin Schmidt Telescope is operated by the California Institute of Technology and Palomar Observatory. The UK Schmidt Telescope was operated by the Royal Observatory Edinburgh, with funding from the UK Science and Engineering Research Council (later the UK Particle Physics and Astronomy Research Council), until 1988 June, and thereafter by the Anglo-Australian Observatory. The blue plates of the southern Sky Atlas and its Equatorial Extension (together known as the SERC-J), as well as the Equatorial Red (ER), and the Second Epoch [red] Survey (SES) were all taken with the UK Schmidt. This work was supported by the National Research Foundation of Korea (NRF) grant funded by the Korean government (MSIT; grant No. 2022R1C1C2004102). 
Y.N. acknowledges support from the Fonds National de la Recherche Scientifique (Belgium), the European Space Agency (ESA) and the Belgian 
Federal Science Policy Office (BELSPO) in the framework of the PRODEX Programme.

%

\vspace{5mm}
\facilities{VLT(FLAMES/GIRAFFE)}


\software{EsoReflex \citep{FRB13}, {\tt IRAF} \citep{T86,T93}, {\tt MIDAS} 
\citep{BGB92}, {\tt NumPy} \citep{HMvdW20}}









\end{document}